# Depth Profile of Optically Recorded Patterns in Light-Sensitive Liquid Crystal Elastomers

Marko Gregorc[1], Boštjan Zalar[1], Valentina Domenici[2], Gabriela Ambrožič[3,4], Irena Drevenšek-Olenik[1,4,5], Martin Fally[6], and Martin Čopič[1,4,5*]

[1]*J. Stefan Institute, Jamova 39, SI 1001 Ljubljana, Slovenia*

[2]*Dipartimento di Chimica e Chimica Industriale, Università degli studi di Pisa, via Risorgimento 35, 56126 Pisa, Italy*

[3]*National Institute of Chemistry, Hajdrihova 19, SI-1000 Ljubljana, Slovenia*

[4]*Faculty of Mathematics and Physics, University of Ljubljana, Jadranska 19, SI 1001 Ljubljana, Slovenia*

[5]*Center of Excellence for Polymer Materials and Technologies, Tehnološki park 24, SI 1000 Ljubljana, Slovenia*

[6]*Faculty of Physics, University of Vienna, Boltzmanngasse 5, A-1090 Wien, Austria*

[*]*Corresponding author: martin.copic@fmf.uni-lj.si*

**Abstract**

We investigated nonlinear absorption and photobleaching processes in a liquid crystal elastomer (LCE) doped with light-sensitive azobenzene moiety. A conventional one-dimensional holographic grating was recorded in the material with the use of two crossed UV laser beams and the angular dependence of the diffraction efficiency in the vicinity of the Bragg peak was analyzed. These measurements gave information on the depth to which *trans* to *cis* isomerisation had progressed into the sample as a function of the UV irradiation time. Using a numerical model that takes into account the propagation of writing beams and rate equations for the local concentration of the absorbing *trans* conformer, we computed the expected spatial distribution of the *trans* and *cis* conformers and the shape of the corresponding Bragg diffraction peak for different irradiation doses. Due to residual absorption of the *cis* conformers the depth of the recording progresses logarithmically with time and is limited by the thermal relaxation from the *cis* to *trans* conformation.

1. Introduction

Liquid crystal elastomers (LCE) show interesting and promising properties arising from the coupling of the liquid crystal order and the elasticity of the crosslinked polymer network. In particular, LCE undergo very large thermal elongation in the direction of the nematic order on cooling from the isotropic to the nematic phase, up to a factor of three and more, and are therefore promising materials for mechanical actuators, e. g. artificial muscles. The deformation is due to the anisotropy of the polymer chain conformation that is induced by the nematic order.

It is well known that a substantial change in the nematic order can also be induced by doping a nematogenic material with photoismerisable molecules like azobenzene and its derivatives that undergo *trans* to *cis* isomerization upon illumination with suitable light[1-6]. The *trans* isomer is of an elongated shape, often nematogenic by itself, while the *cis* isomer is bent and therefore incompatible with nematic ordering. So the light induced *cis* isomers in a nematic host reduce the nematic order and in LCE this also results in large deformations.

There have been a number of experiments on controlling photosensitive LCE samples with light [5, 7, 8], like demonstrations of bending motion. This kind of deformation is the result of inhomogeneous concentration of the *cis* isomer due to the illumination of the sample from one side. The absorption length for UV light that causes the conformation change is of the order of a few micrometers, so the sample contracts only on one side and therefore bends. The absorption process and the resulting bending has been theoretically analyzed by Corbett and Warner[9,10]. These authors show that the linear absorption process is insufficient to explain the observed effects. They include photobleaching to obtain theoretical expressions for the *cis* conformation concentration profiles as a function of the illumination time.

The kinetics of the photoisomerisation process is of fundamental importance for LCE devices controlled by light. In our previous work we reported on holographic patterning of light-sensitive LCE. Due to large obtainable modulation of the refractive indices and possibility of thermal or mechanical control such gratings could find applications in optical devices [11]. In this paper we report on the measurements and modeling of the thickness of the diffraction grating formed holographically with two crossed UV laser beams that produce a modulated concentration of the *cis* isomer. This reduces the degree of nematic order and results in a periodic modulation of the refractive indices. The grating is read out by a visible probe beam. The angular dependence of the diffraction efficiency around the Bragg angle is measured as a function of the illumination time. The width of the Bragg peak is a function of the depth profile of the refractive index modulation and provides information on the effective thickness of the *cis* isomer layer.

## 2. Theory

We consider two laser beams with wavevectors $\mathbf{k}_1$ and $\mathbf{k}_2$ crossing in the sample at an angle $\theta_0$ (Fig. 1). Propagation of light in the sample with some inhomogeneous distribution of the *trans* and *cis* isomers can best be described by using the paraxial wave equation. For this we first write the electric field in the form $E = \sqrt{2\hbar\omega/\bar{\varepsilon}\varepsilon_0 c}\, \psi(x,z)\exp(ikz)$ with $k = \sqrt{\bar{\varepsilon}k_0^2 - (q/2)^2}$, where $\bar{\varepsilon}$ is the average optical dielectric constant of the medium and $(q/2) = (k_0 \sin\theta_0/2)$ is the transverse component of the of the incident wave-vector of each beam. The amplitude function $\psi$ is normalized so that its square is the photon flux density. It is assumed to be a slow function of *z*, so that it satisfies the paraxial wave equation:

$$-2ik\frac{\partial \psi}{\partial z}=\frac{\partial^{2}\psi}{\partial x^{2}}-i\varepsilon''(x,z)k_{0}^{2}\psi \ . \tag{1}$$

The imaginary part of the dielectric function depends on the concentration of the *trans* ans *cis* isomers. It can be expressed with the absorption coefficients

$$\varepsilon''=[\sigma_{t}c_{t}(x,z)+\sigma_{c}c_{c}(x,z)]/k_{0} \ , \tag{2}$$

where $\sigma_{t}$ and $\sigma_{c}$ are the absorption cross-sections for the *trans* and *cis* isomers and $c_{t}(x,z)$ and $c_{c}(x,z)$ are the respective number concentrations. The absorption band of azobenzene derivatives in *trans* conformation has a peak around 350 nm, while the *cis* conformer absorbs in the visible region, but there is some overlap of the absorption bands at the writing laser wavelength. This is important as it limits the maximum depth of the transformed layer. Due to the dependence of the nematic order on the concentration of the *cis* isomer there is also a modulated real part of the dielectric function. While it can be easily included in the computation, we omit it as our measurements of the diffraction efficiency show that it is small and has no effect on the writing process.

The concentrations $c_t$ and $c_c$ at any point in the sample are given by

$$\frac{dc_{t}}{dt}=[-\eta_{t}\sigma_{t}c_{t}+\eta_{c}\sigma_{c}c_{c}]|\psi|^{2}+\frac{c_{c}}{\tau} \ , \tag{3}$$

where $\eta_t$ and $\eta_c$ are constants describing the conversion efficiency from the electronic excited state to the *cis* or *trans* conformation, and $\tau$ is the thermal relaxation time from *cis* to *trans* state. Our measurements were performed at room temperature where this relaxation is of the order of hours, while the writing time for the grating was up to a few minutes, so in the following we will drop the relaxation term. Thermal relaxation, however, is important as it limits the maximum thickness of the transformed layer. We also express the concentrations in the form $c_{t}=c_{t0}n(x,z,t)$ and $c_{c}=c_{t0}(1-n(x,z,t))$, and introduce $\Gamma_{t,c}=\eta_{t,c}\sigma_{t,c}$. Then we have

$$\frac{dn}{dt}=[-\Gamma_{t}n+\Gamma_{c}(1-n)]|\psi|^{2} \ . \tag{4}$$

We can also express the imaginary part of the dielectric function as

$\varepsilon'' = (\mu/k_0^2)[n + (\sigma_c/\sigma_t)(1-n)] \cong \mu n / k_0^2$ with $\mu = \sigma_t c_{t0}$ being the absorption coefficient in the non-bleached regime. The second term in the square parentheses is small and can be omitted in calculations.

We solve equations 1 and 4 numerically by first integrating eq. 1 using Crank – Nicolson scheme for a given *n(x,z,t)* and with the initial condition

$\psi(x,0) = A(e^{iqx/2} + e^{-iqx/2}) = 2A\cos(qx/2)$. Then we get the new *n(x,z,t+Δt)* using eq. 4. We use a grid of 40 points in one period $\Lambda = 2\pi/q$ of the modulation in the *x* direction and 100 points in the *z* direction. The approach using the paraxial wave equation is quite general and can be used with any form of the light field incident on the sample, and with any nonlinear response of the material provided that it is slow compared to the optical frequency.

Before proceeding to the numerical solution we note that for short illumination times, that is when the depth of the recorded grating is small compared to the grating period $\Lambda$, we can neglect the effect of beam coupling and diffraction on the light field propagating into the sample. If in addition we neglect also the transitions from the *cis* back to the *trans* conformation, we can use the analytic solution of ref. [9] that is valid for homogeneous illumination:

$$n(x,z,t) = \frac{1}{e^{t/T(x)-\mu z} - e^{-\mu z} + 1}, \qquad (5)$$

where $T(x) = 1/(\Gamma_t |\psi(x,0)|^2)$. Equation (6) describes a periodic modulation of *n(x,z,t)* in the *x* direction that progresses with a constant rate $1/(\mu T(x))$ into the sample.

The results for the approximate analytic expression (6) and numerical solution of eqs. 1 and 4 are shown in Figure 2. The value of $\varepsilon''$ for pure *trans* isomer was chosen so that for small irradiation dose the experimentally measured decay depth of the grating, equal to $\mu^{-1}$, corresponded with the calculated depth measured as a fraction of $\Lambda$. The measured absorption length is 1.75 μm, that is 1.75 $\Lambda$, in agreement with the published data on azobenzene molar absorption coefficient[12] and the concentration of azobenzene moieties in our sample. This fixed the scaling factor for the $z$ coordinate between the calculation and the experiment. The profile of the concentration of the *trans* isomer gets more and more saturated with time and progresses into the sample. For longer illumination time, due to the effect of beam coupling and diffraction, the numerical result gives considerably different profile from the analytic expression 5. Also, due to the *cis* isomer absorption, the depth of the modulation profile, calculated numerically, increases more and more slowly with longer irradiation, as is clearly evident in Figure 3.

To calculate the resulting diffraction efficiency for the read-out light beam with wavevector $k_r$ incident on the sample at an angle close to the Bragg angle defined by $\sin\theta_B = (k_0/k_r)\sin(\theta_0/2)$, we assume that the magnitude of the nematic order decreases linearly with decreasing concentration of the *trans* isomers. This in turn causes a proportional change in the indices of refraction and components of the dielectric tensor.

The diffraction efficiency the samples in our experiments is much smaller than one, so we can assume that the attenuation of the incoming read-out beam can be neglected and we can use the Born approximation. The dielectric tensor has the form

$$\varepsilon(x,z) = \varepsilon_0 + \varepsilon_1(z)\cos qx + \varepsilon_2(z)\cos 2qx + \ldots \quad . \tag{6}$$

The modulated terms in this expansion are proportional to the modulation of the *trans* conformer in the sample. Then the amplitude of the diffracted beam in the vicinity of the Bragg angle is given by

$$E_{diff}(\theta) = \frac{iE_0 k_r^2}{2k_z} \int_0^L \varepsilon_1(z) e^{ipz} dz \quad , \tag{7}$$

where $p = (k_r q/k_z)(\theta - \theta_B)\cos\theta_B$ and $k_z = \sqrt{\varepsilon_0 k_r^2 - q^2/4}$. So the diffracted amplitude is simply given by the Fourier transform of $\varepsilon_1(z)$.

Figure 3 shows $\varepsilon_1(z)$ obtained from the numerical calculation of $n(x,z,t)$ for several values of the irradiation time. At a given depth $\varepsilon_1$ first increases then, due to the saturation of the *cis* conformer, decreases. The rate of progression of the maximum in $\varepsilon_1$ with time decreases for irradiation time $t$ long compared to $T(0) = 1/(\Gamma_t |\psi(0,0)|^2)$ due to the absorption of the *cis* isomer and transitions back to the *trans* conformation. In this regime, as long as $t$ is smaller than the thermal relaxation time $\tau$ from cis to trans conformation, the depth of the profile increases approximately logarithmically with $t$. To understand this and to estimate the maximum depth of the transformed layer, we can assume homogeneous illumination. Then the stationary (saturated) solution of eq. 3 is

$$n_s = \frac{\Gamma_c |\psi(z)|^2 + 1/\tau}{(\Gamma_t + \Gamma_c)|\psi(z)|^2 + 1/\tau} \qquad (8)$$

As long as $t \ll \tau$, the stationary ratio of the *cis* to *trans* isomer is independent of the light intensity, which decays approximately exponentially with an absorption coefficient $\mu_s = \mu(n_s + \Gamma_c/\Gamma_t(1-n_s)) \approx 2\mu\Gamma_c/\Gamma_t$. The time to get to the stationary state is of the order of $1/(\Gamma_t |\psi(z)|^2)$, so the depth of the saturated layer is a logarithmic function of time. The maximum depth is reached when $1/(\Gamma_t |\psi(z)|^2) \approx \tau$ and its value is approximately

$$d_{max} = \Gamma_t/(2\Gamma_c \mu) \ln(\Gamma_t |\psi(0)|^2 \tau).$$

The diffraction efficiency around Bragg angle, computed from equation (7), for several values of the irradiation times is shown in Figure 4. Clearly the width of the Bragg peak becomes narrower for longer irradiation times. The effective depth of the grating can be reasonably estimated from the half-width of the diffraction peak: $L_{eff} = k_z/(k_r q \cos\theta_B \Delta\theta_{1/2})$. Also, for longer illumination times $\varepsilon_1(z)$ is somewhat box-shaped and its Fourier *trans*form shows a small side-lobe. So measurements of the Bragg peak width allow us to determine the depth of the modulation of the *trans* and *cis* conformation concentration of azobenzene. In principle, it should also be possible to obtain the functional form of $\varepsilon_1(z)$, but in practice this is probably rather difficult due to measurement noise and sample imperfections.

### 3. Experimental results

Macroscopically aligned (monodomain) side-chain LCE films with a thickness of 150 μm were prepared according to the two-step "Finkelmann crosslinking procedure". The polymer back-bone is based on a commercial hydroxymethyl-polysiloxane, which is crosslinked by 1,4-bis (undec-10-en-1-yloxy) benzene used as crosslinker unit. The side-chain moieties are composed of usual rod-like mesogens (4-methoxyphenyl 4-(but-3-en-1-yloxy) benzoate) and light-sensitive azomesogens (1-(4-(hex-5-enyloxy)phenyl)-2-(4-methoxyphenyl) diazene) in the ratio of 9:1 [13]. The transition to the isotropic state is at 80°C.

The diffraction gratings were written with two beams with wave-length 351 nm obtained from an argon ion laser. The power of each beam was 10 mW. The angle between the beams was $20^0$, so that the period of the resulting grating was 1 μm. The nematic director was perpendicular to the plane of the beams, and the polarization of both writing beams was parallel to the director, i.e., the beams were extraordinarily polarized. The diffraction efficiency of the grating was measured by a low power HeNe laser beam at 632.8 nm that was also polarized parallel to the director. The sample was rotated around the position fulfilling Bragg condition to obtain the width of the Bragg peak. The temperature for all measurements was 35°C. At this temperature the back relaxation from *cis* to *trans* conformation is around 1 hour. After each measurement the sample was heated to the isotropic phase and cooled back to T=35°C to ensure complete erasure of the previous grating before recording a new one.

Figure 5 shows the angular dependence (rocking curve) of the diffraction efficiency around the Bragg angle for several UV illumination times. The Bragg peak clearly narrows for longer times. The peaks are fitted with a lorentzian function to obtain a measure of the peak width. In one or two cases the Bragg peak shows a side-lobe as expected from numerical simulation, but in most cases the side-lobe was not visible and Lorentzian shape was a reasonable approximation. The absence of the side-lobes is most probably due to imperfections and inhomogeneities of the sample.

Figure 6 shows the effective depth of the grating $L_{\text{eff}}$ calculated from measured inverse width of the Bragg peak as a function of the illumination time. Also shown as a full line is the theoretical result for $L_{\text{eff}}$ obtained with the following values of the parameters: $\mu = \sigma_t c_{t0} = 1.75\ \mu\text{m}$, $\sigma_c/\sigma_t = \Gamma_c/\Gamma_t = 0.068$, and $T(0) = 1/(\Gamma_t |\psi(0,0)|^2) = 1.4$ s. The depth of the recorded grating goes from approximately 1.7 $\mu$m to 20 $\mu$m. Logarithmic dependence for irradiation times longer than about 10 s is clearly visible. The theoretical limiting depth would be reached in about 3000 s and is about 100 $\mu$m. In our experiments illumination times longer than 1000 s did not lead to a markedly narrower Bragg peak. The peak is probably broadened by variations of the grating period that can change due to the mechanical strain that is induced by the decrease in the nematic order. This effect is not as large as one would expect because through most of the sample the *trans* concentration is not changed and hence the sample is not allowed to contract, so that the illuminated layer is effectively mechanically quenched.

It is very interesting to compare the obtained values of the parameters with the known values for azobenzene in low molar weight solvents like toluene. Our value for the *trans* isomer absorption coefficient is well in accord with the published values [12], as already remarked above. It is also well known that the transitions from the excited electronic level to either *trans* or *cis* conformation occur on the time-scale of picoseconds [14,15,16]. In low molecular weight solvents the quantum efficiency $\eta_t$ for the transition from the electronic excited state to the *cis* conformation is between 0.2 and 0.3. The molar absorption coefficient is 4 x $10^4$ cm$^{-1}$, so that $\sigma_t = 10^{-16}$ cm$^2$. In our experiment the average incoming light flux density was 10 mW/cm² giving the peak photon flux density $|\psi(0,0)|^2 = 8\times10^{16}/\text{cm}^2$. The characteristic time $T(0) = 1/(\eta_t \sigma_t |\psi(0,0)|^2)$ with $\eta_t = 0.2$ would then be 1 s, not much smaller than the value 1.4 s that we get in LCE. This result is a actually little surprising, as one would expect that the higher viscosity of the polymer network and the constrain of the nematic director field that favors *trans* conformation would more strongly decrease the quantum efficiency for the transition to the *cis* state.

4.  **Conclusion**

Holographically recorded diffraction gratings in photosensitive LCE are potentially useful in optical beam steering and similar devices as the diffraction properties can be manipulated by mechanical strain or temperature. Besides this, our analysis shows that by such gratings it is also possible to obtain information on the distribution of the *cis* conformer in a sample illuminated by UV light by measuring the angular width of the Bragg diffraction peak. From our results it follows that the writing process strongly depends on photo-bleaching of the *trans* conformer of the photosensitive moieties in LCE. In addition to this, due to the absorption of the *cis* isomer the depth of the recorded grating progresses logarithmically with time. For long exposure times the thermally driven transitions from the *cis* to *trans* conformation limits the maximum depth of the recording.

5.  **Acknowledgements**

This work was supported by bilateral project BI-AT/11-12-015 and Slovenian research programme P1-0192 - Light and Matter. The authors acknowledge the financial support from the Ministry of Higher Education, Science and Technology of the Republic of Slovenia through the contract No. 3211-10-000057 (Center of Excellence for Polymer Materials and Technologies).

FIGURE CAPTIONS

Figure 1. The experimental geometry used in writing the diffraction gratings.

Figure 2. Calculated distributions of the *trans* conformer. a) and b) show the analytic expression (5) for $t = 1$ and $t = 20$, and c) and d) show the results of numerical calculation for $t = 1$ and $t = 20$. Time is measured in units of $T(0)$ and $z$ in units of $\Lambda$.

Figure 3. The calculated dependence of the first Fourier component of the optical dielectric constant $\varepsilon_1$ on $z$ for illumination times $t = 1$ (full curve), 5 (long dashes), 20 (short dashes), 40 (dot-dashed), and 60 (dotted). Time is measured in units of $T(0)$ and $z$ in units of $\Lambda$.

Figure 4. Calculated Bragg diffraction peaks, scaled to unit height, for illumination times $t = 1$ (full curve), 5 (long dashes), 20 (short dashes), 40 (dot-dashed), and 60 (dotted). Time is measured in units of $T(0)$ and $\theta - \theta_B$ is given in radians.

Figure 5. Measured angular dependence of the Bragg peak with fits of Lorentzian profile for different illumination times.

Figure 6. The measured depth of the recorded grating, obtained from the half-widths of the Bragg peak, as a function of the illumination time. The average UV light flux density was 10 mW/cm$^2$. The full line shows the result of the numerical model obtained with the parameters described in the text.

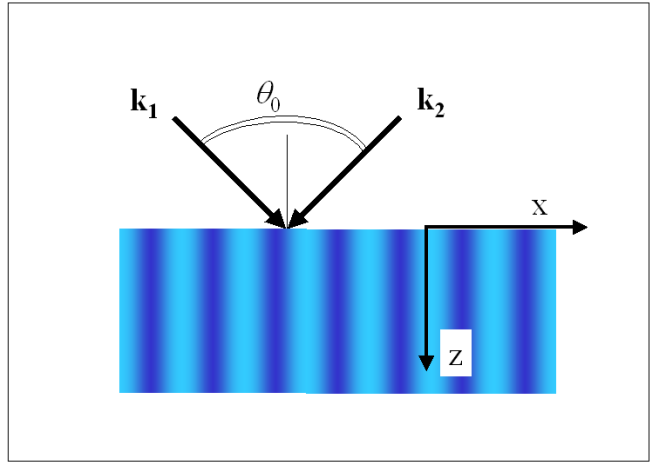

Figure 1.

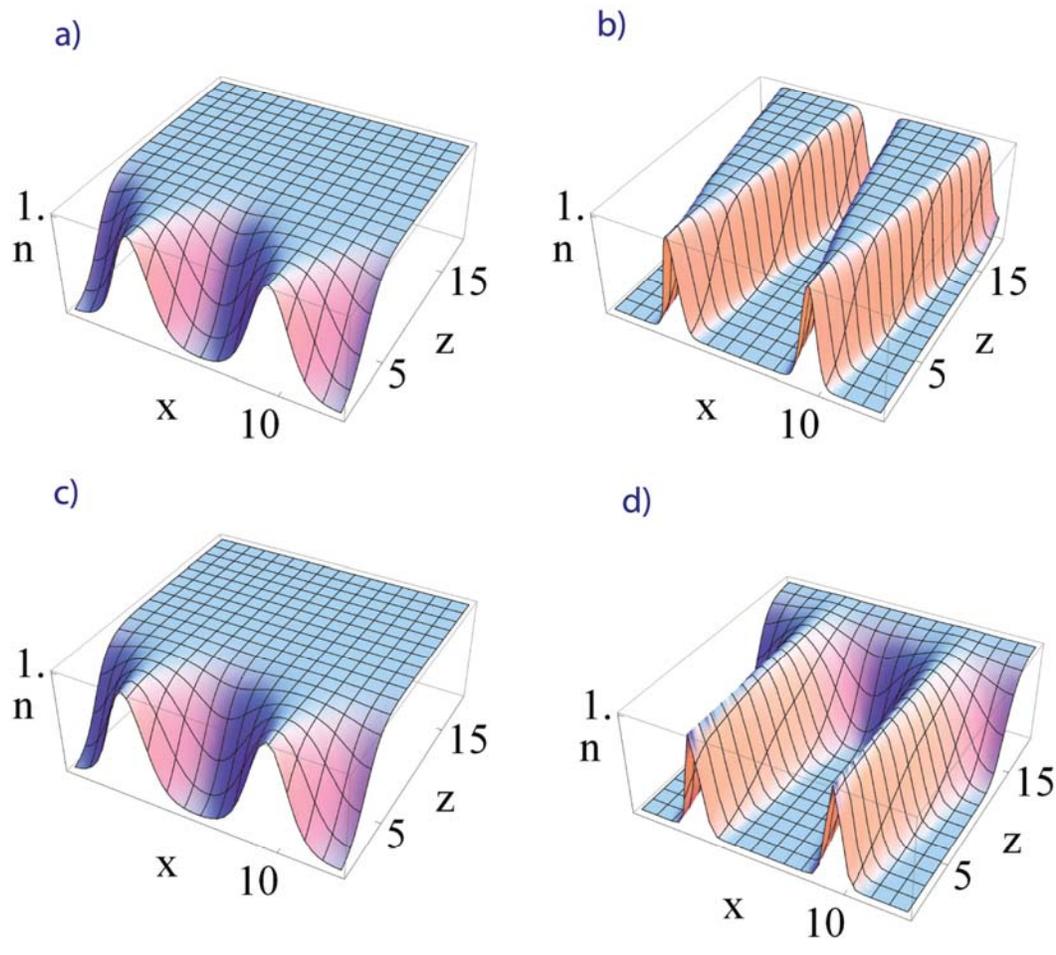

Figure 2.

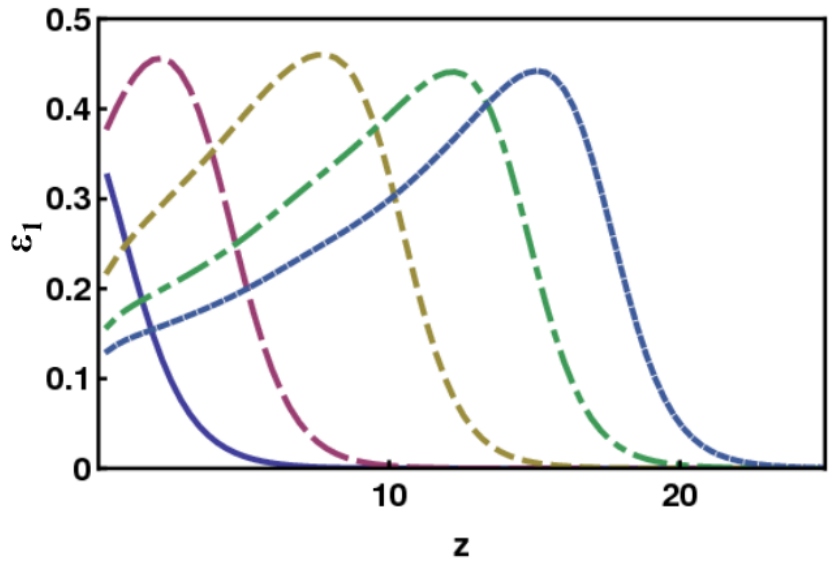

Figure 3.

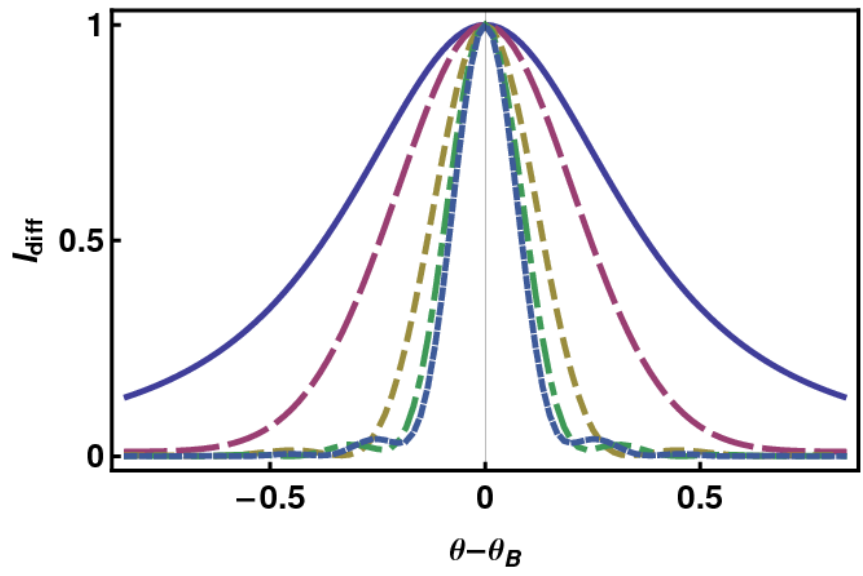

Figure 4.

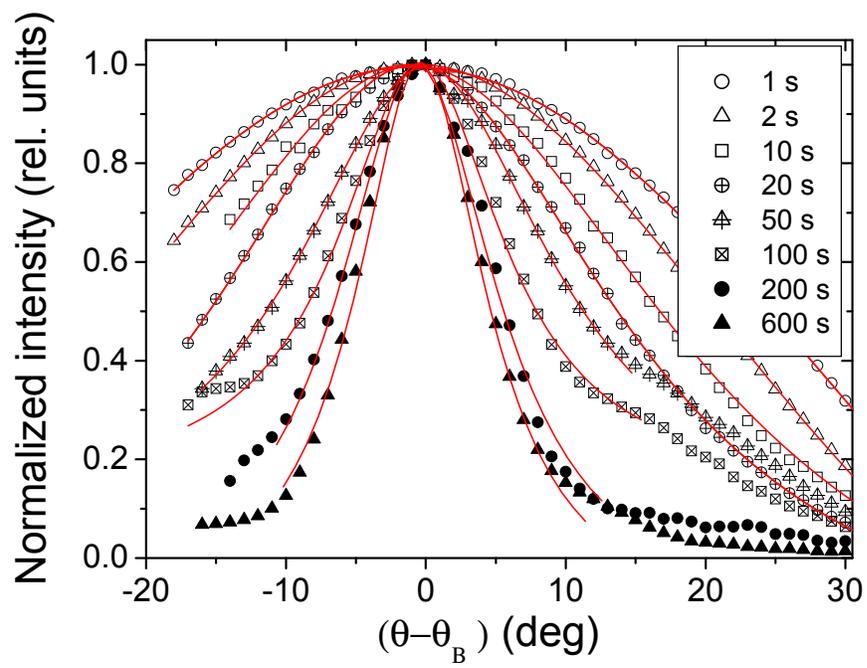

Figure 5.

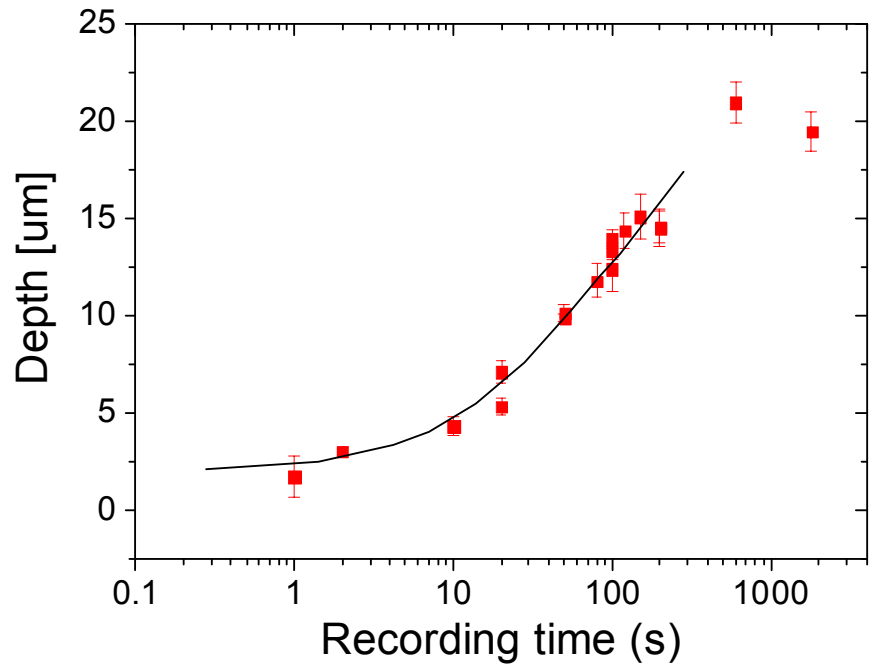

Figure 6.